# Continuous Speech Recognition Based on Deterministic Finite Automata Machine using Utterance and Pitch Verification.

M.THARUN PRASATH
EMAIL ID: tharunprasathcse@gmail.com, MOBILE NO: 9688949919.

*Abstract*—This paper introduces a set of acoustic modeling techniques for utterance verification (UV) based continuous speech recognition (CSR). Utterance verification in this work implies the ability to determine when portions of a hypothesized word string correspond to incorrectly decoded vocabulary words or out-of-vocabulary words that may appear in an utterance. This capability is implemented here as a likelihood ratio (LR). There are two UV techniques that are presented here. The first is voice verification along with the vocabulary testing, at the same time the parameters for UV models are generated based on the optimization criterion which is directly related to the LR measure. The second technique is a pitch recognition based on weighted finite-state transducers. These transducers provide a common and natural representation for major components of speech recognition systems, including hidden Markov models (HMMs), context-dependency models, pronunciation dictionaries, statistical grammars, and word or phone lattices. The finite state machine processes the acoustic parameters of UV models. The results of an experimental study presented in the paper shows that LR based parameter estimation results in a significant improvement in UV performance for this task. The study also found that the use of the LR based weighted finite-state transducers along with the UV, can provide as much as an 11% improvement in UV performance when compared to existing UV procedures. Finally, it was also found that the performance of the finite state machine was highly dependent on the use of the LR criterion in training acoustic models. Several observations are made in the paper concerning the formation of confidence measures for UV and the interaction of these techniques with statistical language models.

## 1. INTRODUCTION

In Continuous speech recognition applications it is often necessary to provide a mechanism for verifying the accuracy of portions of recognition hypotheses. This paper describes utterance verification (UV) procedures for weighted finite state automata model based on a Likelihood Ratio (LR) criterion. Utterance verification is most often considered as a hypothesis testing problem. Existing techniques rely on a speech recognizer to produce a hypothesized word or string of words along with hypothesized word boundaries obtained through Viterbi segmentation of the utterance. A measure of confidence is then assigned to the hypothesized string, and the hypothesized word labels are accepted or rejected by comparing the confidence measure to a decision threshold. LR based Weighted Finite State Model (WFSM) process the pitch parameters with respect to a source hypothesis WFS model which is already obtained. A finite-state transducer is a finite automaton whose state transitions are labeled with both input and output symbols. A weighted transducer puts weights on transitions in addition to the input and output symbols. Weights may encode probabilities, durations, penalties, or any other quantity that accumulates along paths to compute the overall weight of mapping an input string to an out-put string. Weighted transducers are thus a natural choice to represent the probabilistic finite-state models prevalent in speech processing.

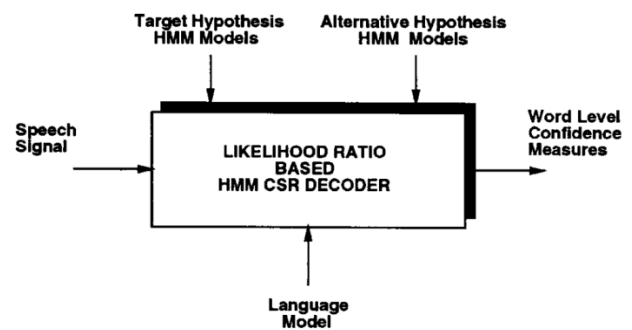

## 2. OVERVIEW

We start with an informal overview of weighted automata and transducers, outlines of some of the key algorithms that support the utterance and weighted finite automata verifications described in this paper – composition, determination, and minimization, and dynamic verification.

### 2.1. WEIGHTED ACCEPTORS

Weighted finite automata (or weighted acceptors) are used widely in continuous speech recognition (CSR). Figure 1 gives simple, familiar examples of weighted automata as used in CSR. The automaton in Figure 1(a) is a toy finite-state language model. The legal word strings are specified by the words along each complete path, and their probabilities by the product of the corresponding transition probabilities. The automaton in Figure 1(b) gives the possible pronunciations of one word, data, used in the language model. Each legal pronunciation is the phone strings along a complete path, and its probability is given by the product of the corresponding transition probabilities. Finally, the automaton in Figure 1(c) encodes a typical left-to-right, three-distribution-HMM structure for one micro phone, with the labels along a complete path specifying legal strings of acoustic distributions for that micro phone.

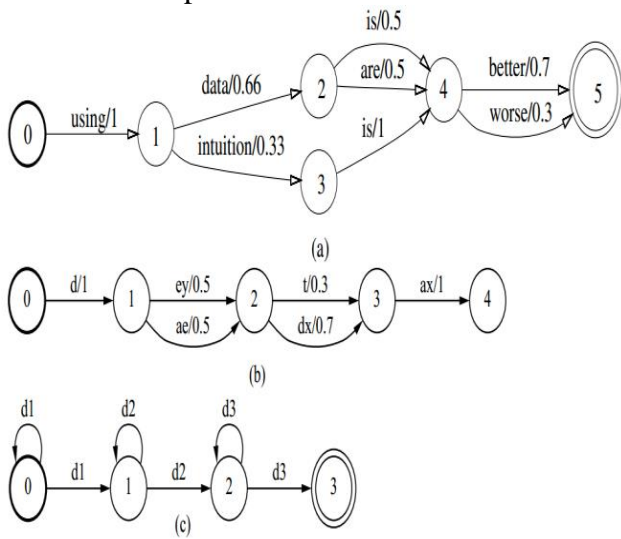

Figure 1: Weighted finite-state acceptor examples.

### 2.2. WEIGHTED TRANSDUCERS

Our approach uses finite-state transducers, rather than acceptors, to represent the n-gram grammars, pronunciation dictionaries, context-dependency specifications, HMM topology, word, phone or HMM segmentations, lattices and n-best output lists encountered in CSR. The transducer representation provides general methods for combining models and optimizing them, leading to both simple and flexible CSR designs. A weighted finite-state transducer (WFST) is quite similar to a weighted acceptor except that it has an input label, an output label and a weight on each of its transitions. The examples in Figure 2 encode (a superset of) the information in the WFSAs of figure 1(a)-(b) as WFSTs. Figure 2(a) represents the same language model as Figure 1(a) by giving each transition identical input and output labels. This adds no new information, but is a convenient way we use often to treat acceptors and transducers uniformly.

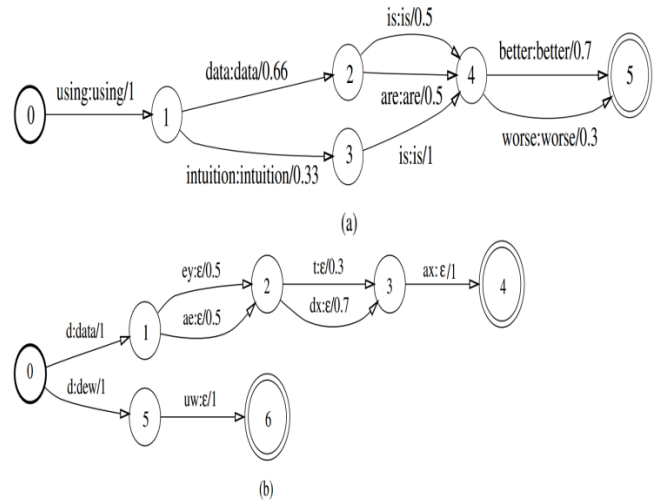

Figure 2: Weighted finite-state transducer examples.

### 2.3. COMPOSITION

Composition is the transducer operation for combining different levels of representation. For in-stance, a pronunciation lexicon can be composed with a word-level grammar to produce a phone-to-word transducer whose word strings are restricted to the grammar. A variety of CSR transducer combination techniques, both context-independent and context-dependent, are conveniently and efficiently implemented with composition.

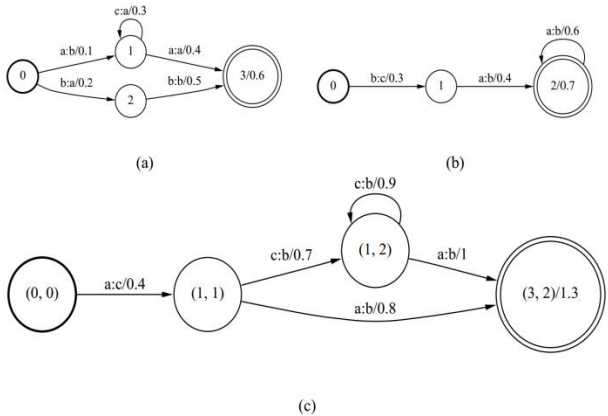

Figure 3: Example of transducer composition.

## 2.4. DETERMINATION

In a deterministic automaton, each state has at most one transition with any given input label and there is no input -labels. Figure 4(a) gives an example of a non-deterministic weighted acceptor: at state 0, for instance, there are two transitions with the same label a. The automaton in Figure 4(b), on the other hand, is deterministic. The key advantage of a deterministic automaton over equivalent nondeterministic ones is its ir-redundancy: it contains at most one path matching any given input string, thereby reducing the time and space needed to process the string. This is particularly important in ASR due to pronunciation lexicon redundancy in large vocabulary tasks.

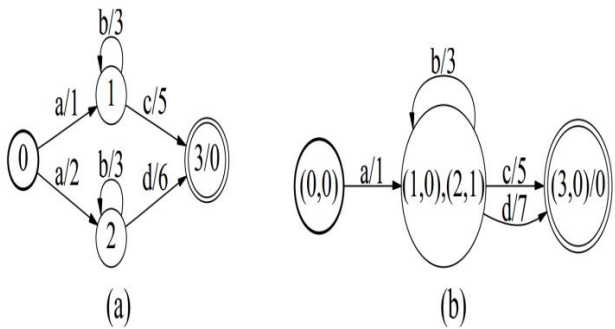

Figure 4: Determination of weighted automata. (a) Weighted automaton over the tropical semi ring A. (b) Equivalent weighted automaton B obtained by determination of A.

## 2.5. MINIMIZATION

Given a deterministic automaton, we can reduce its size by minimization, which can save both space and time. Any deterministic un-weighted automaton can be minimized using classical algorithms [Ahoet al., 1974, Revuz, 1992]. In the same way, any deterministic weighted automaton A can be minimized using our minimization algorithm, which extends the classical algorithm [Mohri, 1997]. The resulting weighted automaton B is equivalent to the automaton A, and has the least number of states and the least number of transitions among all deterministic weighted automata equivalent to A. As we will see in Section 3.5, weighted minimization is quite efficient, indeed as efficient as classical deterministic finite automata (DFA) minimization.

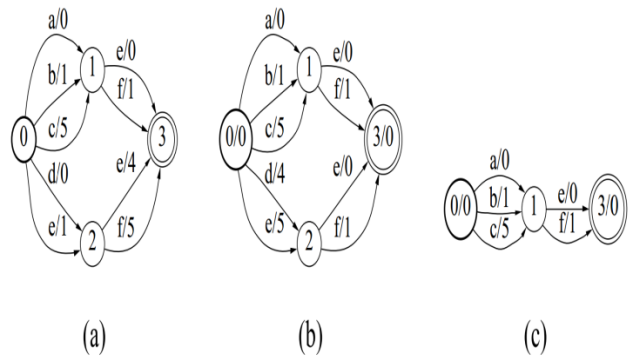

Figure 5: Weight pushing and minimization. (a) Deterministic weighted automaton A. (b) Equivalent weighted automaton B obtained by weight pushing in the tropical semi ring. (c) Minimal weighted automaton equivalent to A.

## 2.6. SPEECH RECOGNITION TRANSDUCERS

As an illustration of these methods applied to speech recognition, we describe how to construct a single, statically-compiled and optimized recognition transducer that maps from context-dependent phones to words. This is an attractive choice for tasks that have fixed acoustic, lexical, and grammatical models since the static transducer can be searched simply and efficiently with no recognition-time overhead for model combination and optimization.

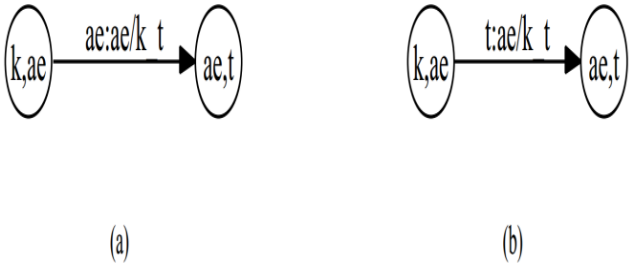

Figure 6: Context-dependent tri-phone transducer transition: (a) non-deterministic, (b) deterministic.

# 3. ALGORITHMS

We now describe in detail the weighted automata and transducer algorithms introduced informally in Section 2 that are relevant to the design of speech recognition systems. We start with definitions and notation used in specifying and describing the algorithms.

```
WEIGHTED-COMPOSITION(T_1, T_2)
1   Q ← I_1 × I_2
2   S ← I_1 × I_2
3   while S ≠ ∅ do
4       (q_1, q_2) ← HEAD(S)
5       DEQUEUE(S)
6       if (q_1, q_2) ∈ I_1 × I_2 then
7           I ← I ∪ {(q_1, q_2)}
8           λ(q_1, q_2) ← λ_1(q_1) ⊗ λ_2(q_2)
9       if (q_1, q_2) ∈ F_1 × F_2 then
10          F ← F ∪ {(q_1, q_2)}
11          ρ(q_1, q_2) ← ρ_1(q_1) ⊗ ρ_2(q_2)
12      for each (e_1, e_2) ∈ E[q_1] × E[q_2] such that o[e_1] = i[e_2] do
13          if (n[e_1], n[e_2]) ∉ Q then
14              Q ← Q ∪ {(n[e_1], n[e_2])}
15              ENQUEUE(S, (n[e_1], n[e_2]))
16          E ← E ∪ {((q_1, q_2), i[e_1], o[e_2], w[e_1] ⊗ w[e_2], (n[e_1], n[e_2]))}
17  return T
```

Figure 7: Pseudo code of the composition algorithm.

## 3.1. DETERMINATION

We now describe the generic determination algorithm for weighted automata that we used informally when working through the example in Section 2.4. This algorithm is a generalization of the classical subset construction for NFAs (un-weighted nondeterministic finite automata). The determination of un-weighted or weighted finite-state transducers can both be viewed as special instances of the generic algorithm presented here but, for simplicity, we will focus on the weighted acceptor case. A weighted automaton is deterministic (also known as sub sequential) if it has a unique initial state and if no two transitions leaving any state share the same input label. The determination algorithm we now present applies to weighted automata over a cancellative weakly left-divisible semi ring that satisfies a mild technical condition. Figure 10 gives pseudo code for the algorithm.

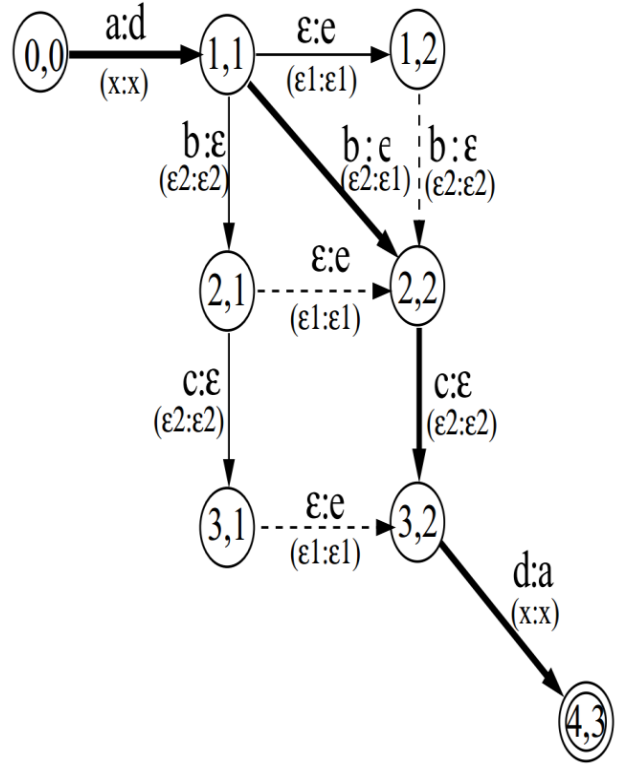

*Figure 8:* Redundant ε-paths. A straightforward generalization of the -free case could generate all the paths from (1, 1) to (3, 2) when composing the two simple transducers on the right-hand side.

**Theorem 1** Let A be a weighted automaton over the tropical semi ring. If A has the twins property, then A is determinable.

**Theorem 2** Let A is a trim unambiguous weighted automaton over the tropical semi ring. Then A is determinable iff it has the twins property.

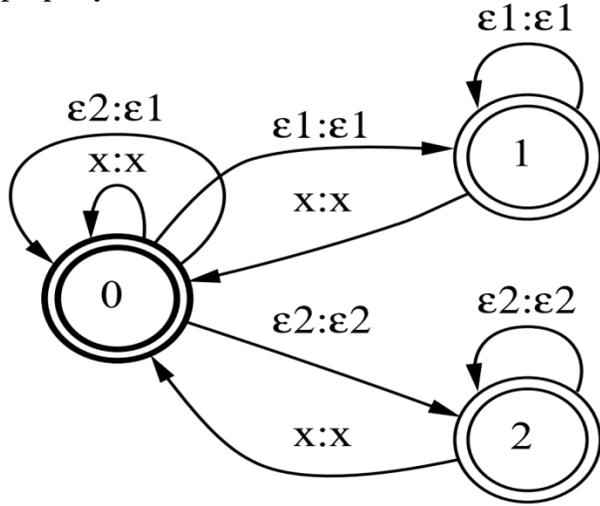

Figure 9: Filter for composition F.

```
WEIGHTED-DETERMINIZATION(A)
1  i' ← {(i, λ(i)) : i ∈ I}
2  λ'(i') ← 1̄
3  S ← {i'}
4  while S ≠ ∅ do
5      p' ← HEAD(S)
6      DEQUEUE(S)
7      for each x ∈ i[E[Q[p']]] do
8          w' ← ⊕{v ⊗ w : (p, v) ∈ p', (p, x, w, q) ∈ E}
9          q' ← {(q, ⊕{w'^{-1} ⊗ (v ⊗ w) : (p, v) ∈ p', (p, x, w, q) ∈ E}) :
                 q = n[e], i[e] = x, e ∈ E[Q[p']]}
10         E' ← E' ∪ {(p', x, w', q')}
11         if q' ∉ Q' then
12             Q' ← Q' ∪ {q'}
13             if Q[q'] ∩ F ≠ ∅ then
14                 F' ← F' ∪ {q'}
15             ρ'(q') ← ⊕{v ⊗ ρ(q) : (q, v) ∈ q', q ∈ F}
16             ENQUEUE(S, q')
17 return T'
```

Figure 10: Pseudo code of the weighted determination algorithm [Mohri, 1997].

The pre-determination algorithm can be used to make determinable an arbitrary weighted transducer over the tropical semi ring by inserting transitions labeled with special symbols [Allauzen and Mohri, 2004]. The algorithm makes use of a general twins property [Allauzen and Mohri, 2003] to insert new transitions when needed to guarantee that the resulting transducer has the twins property and thus is determinable.

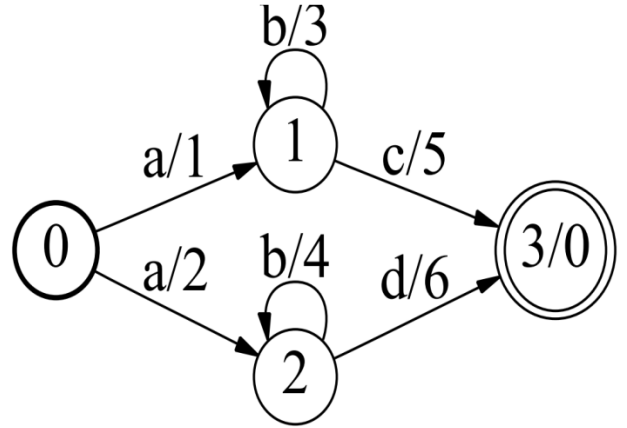

Figure 11: Non-determinable weighted automaton over the tropical semi ring. States 1 and 2 are non-twin siblings.

### 3.2. WEIGHT PUSHING

As discussed in Section 2.5, weight pushing is necessary in weighted minimization, and is also very useful to improve search. Weight pushing can also be used to test the equivalence of two weighted automata. Weight pushing is possible because the choice of the distribution of the total weight along each successful path of a weighted automaton does not affect the total weight of each successful path, and therefore preserves the definition of the automaton as a weighted set (weighted relation for a transducer).

**Theorem 3** Let A be a deterministic weighted automaton over a semi ring K. Assume that the conditions of application of the weight pushing algorithm hold. Then the execution of the following steps:

1. Weight pushing.
2. (Un-weighted) automata minimization, lead to a minimal weighted automaton equivalent to A.

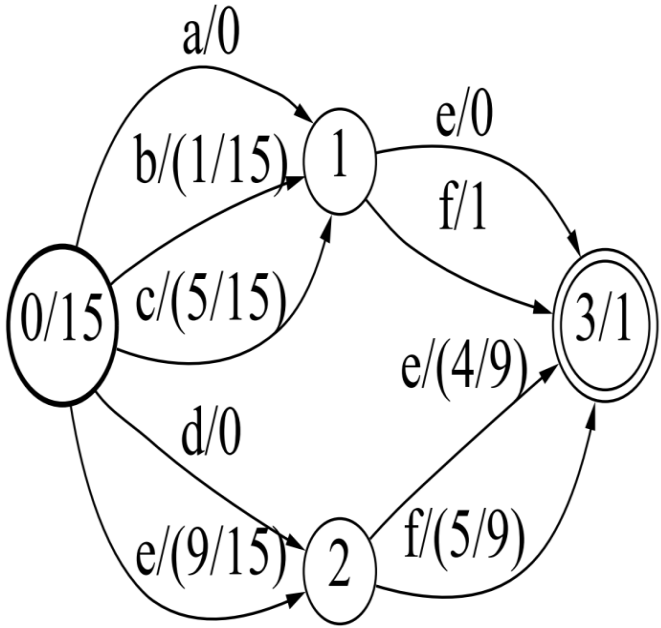

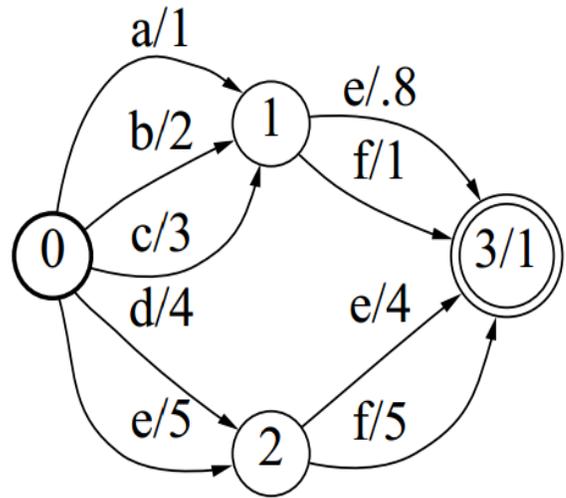

Figure 13: Minimization of weighted automata.

Figure 12: Weighted automaton C obtained from A of Figure 5(a) by weight pushing in the probability semi ring.

## 3.3. MINIMIZATION

Finally, we discuss in more detail the minimization algorithm introduced in Section 2.5. A deterministic weighted automaton is said to be minimal if there is no other deterministic weighted automaton with a smaller number of states that represents the same mapping from strings to weights. It can be shown that the minimal deterministic weighted automaton has also the minimal number of transitions among all equivalent deterministic weighted automata [Mohri, 1997].

Two states of a deterministic weighted automaton are said to be equivalent if exactly the same set of strings label the paths from these states to a final state, and the total weight of the paths for each string, including the final weight of the final state, is the same. Thus, two equivalent states of a deterministic weighted automaton can be merged without affecting the function realized by that automaton. A weighted automaton is minimal when it is not possible to create two distinct equivalent states after any pushing of the weights along its paths.

## 4. CONCLUSION

We presented an overview of weighted finite-state transducer for UV based continuous speech recognition. The methods are quite general, and can also be applied in other areas of speech and language processing, including information extraction, speech synthesis, and other pattern matching and string processing applications, just to mention some of the most active application areas. By adding additional pitch verification by WFSM along with the UV based CSR we could improve 11% accuracy. As we uses the pitch verification parameters from the LR generated for the UV, there is no additional module is required. As the pitch verification is software module attached to the UV based CSR, we could implement this two phase verification at a lower cost.

## REFERENCES

[1] Alfred V. Aho, John E. Hopcroft, and Jeffrey D. Ull-man. The design and analysis of computer algorithms. Addison Wesley, Reading, MA, 1974.

[2] Alfred V. Aho, Ravi Sethi, and Jeffrey D. Ullman. Compilers, Principles, Techniques and Tools. Ad-dison Wesley, Reading, MA, 1986.

[3] Cyril Allauzen and Mehryar Mohri. An Optimal Pre-Determination Algorithm for Weighted Transducers. Theoretical

Computer Science, 328(1-2): 3–18, November 2004.
[4] Cyril Allauzen and Mehryar Mohri. Efficient Algorithms for Testing the Twins Property. Journal of Automata, Languages and Combinatorics, 8(2): 117–144, 2003.
[5] Cyril Allauzen, Mehryar Mohri, and Michael Riley. Statistical Modeling for Unit Selection in Speech Synthesis. In 42nd Meeting of the Association for Computational Linguistics (ACL 2004), Proceedings of the Conference, Barcelona, Spain, July 2004a.
[6] Cyril Allauzen, Mehryar Mohri, Brian Roark, and Michael Riley. A Generalized Construction of Integrated Speech Recognition Transducers. In Proceedings of the International Conference on Acoustics, Speech, and Signal Processing (ICASSP 2004), Montreal, Canada, May 2004b.
[7] Jean Berstel. Transductions and Context-Free Languages. Teubner Studienbucher, Stuttgart, 1979.
[8] Jean Berstel and Christophe Reutenauer. Rational Series and Their Languages. Springer-Verlag, Berlin-New York, 1988.
[9] Jack W. Carlyle and Azaria Paz. Realizations by Stochastic Finite Automaton. Journal of Computer and System Sciences, 5:26–40, 1971.
[10] Maxime Crochemore and Wojciech Rytter. Text Algorithms. Oxford University Press, 1994.
[11] Karel Culik II and Jarkko Kari. Digital Images and Formal Languages. In Grzegorz Rozenberg and Arto Salomaa, editors, Handbook of Formal Languages, pages 599–616. Springer, 1997.
[12] Samuel Eilenberg. Automata, Languages and Ma-chines, volume A-B. Academic Press, 1974-1976.
[13] John E. Hopcroft and Jeffrey D. Ullman. Introduction to Automata Theory, Languages, and Computation. Addison Wesley, Reading, MA, 1979.
[14] Arto Salomaa and Matti Soittola. Automata-Theoretic Aspects of Formal Power Series. Springer-Verlag, New York, 1978.
[15] Muret Saraclar, Michael Riley, Enrico Bocchieri, and Vincent Goffin. Towards automatic closed captioning: Low latency real time broadcast news transcription. In Proceedings of the International Conference on Spoken Language Processing (IC-SLP'02), 2002.